# Towards the Interplanetary Internet: An IoT Perspective

Carles Gomez[1], Jon Crowcroft[2]
[1]*Universitat Politècnica de Catalunya*, [2]*University of Cambridge*

***Abstract*—Public administrations and private companies have announced plans to deploy networking infrastructure to support future robotic and human presence on or near space targets, such as the Moon and Mars. While using an IP protocol stack for deep-space communication had been neglected, recent events have motivated the reconsideration of IP to enable the Interplanetary Internet. This new paradigm facilitates the integration of IP-based Internet of Things (IoT) protocols for deep-space environments. This paper illustrates the similarities between deep-space and IoT scenarios, presents related IETF standardization work, and discusses opportunities and future directions for IP-based IoT protocols in the Interplanetary Internet.**

## I. Introduction

Deep-space communication has been used in support of space exploration for decades. The first ever interplanetary communication occurred in 1962, when NASA's Mariner 2 probe approached Venus, collected scientific data about its surface and atmosphere, and sent it to Earth at 8 bit/s. In the following years, an intense space activity took place focusing on the inner solar system. As the Cold War declined, space missions decreased substantially. However, in recent years, the space sector has regained significant momentum (Fig. 1), with private companies and public administrations aiming to capture new space markets expected to reach trillions of dollars [1], and announcing plans for crewed expeditions to the Moon and Mars [2].

In the same decade that witnessed the dawn of deep-space exploration, another crucial milestone was attained: the ARPAnet, the first embryo of the Internet, was created. Subsequently, the network expanded massively, ultimately becoming the Internet of Things (IoT), and reaching a fundamental role in our society.

In the early 2000s, anticipating the communication needs in deep-space environments, researchers analyzed whether the core Internet communication protocols, i.e., the TCP/IP protocol stack, would be suitable to enable the Interplanetary Internet. They concluded that "Internet protocols do not work well" in such an environment, characterized by challenging features, such as extreme delay and connectivity disruption [3]. As a result, they designed the Delay-Tolerant Networking Architecture, based on a message-oriented overlay between the application layer and the transport layer called the Bundle Protocol (BP) [4].

For a quarter of a century, the IP-based protocol stack has been neglected as a potential candidate for end-to-end communication in deep space. However, recent plans to deploy IP-based networking infrastructure on or near space targets (e.g., Moon, Mars, etc.), the emergence of new IP-based transport protocols (e.g., QUIC), and the revision of old assumptions have motivated reevaluating the use of an IP-based protocol stack for end-to-end communication in the Interplanetary Internet [5]. Capturing this new assessment, the Taking IP to Other Planets (TIPTOP) working group (WG) was recently established by the IETF to provide guidance on using an IP-based protocol stack for deep-space communication.

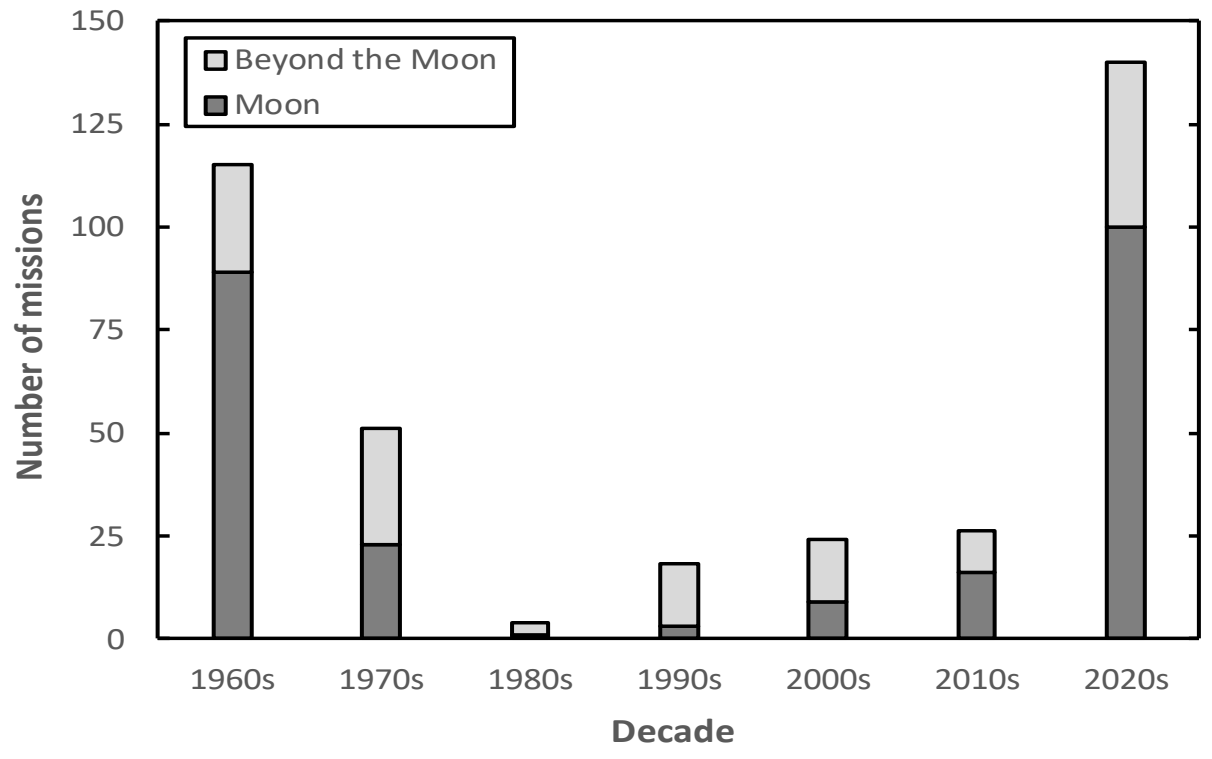

Fig. 1. Number of missions to the Moon and beyond the Moon over time. The data for the 2020s comprises already launched and planned missions.

On the other hand, for the last two decades, the IETF has also been developing IP-based communication protocols suitable for the IoT, covering areas such as packet header compression, packet fragmentation, reliability, congestion control, application support, security, and management. Remarkably, deep-space environments and IoT scenarios share similar networking characteristics. Accordingly, there are significant opportunities for IoT protocols to be used in deep-space environments. However, such protocols also face challenges and require adaptation for proper operation in deep space.

In this paper, we depict the similarities between deep-space and IoT scenarios, we present related IETF standardization work, and we discuss opportunities and future directions for IP-based IoT protocols in the Interplanetary Internet. The remainder of the paper is organized as follows. Section II illustrates deep-space scenarios and IoT use cases therein. Section III analyzes the networking characteristics of deep-space and IoT environments. Section IV briefly overviews the BP-based and the IP-based architectures for deep space, and introduces an IoT-based architecture for such a scenario. Section V describes IoT protocols in a deep-space context, whereas Section VI discusses their adaptation for such environments, including future directions. Section VII concludes the paper.

## II. Deep Space: Definitions, Mission Environments, and IoT Use Cases

This section discusses the definition of the term *deep space*, it illustrates current and future deep-space mission scenarios, and it overviews deep-space IoT use cases.

### A. Definitions

*Deep space* has different meanings depending on the considered context. In astronomy, deep space refers to objects outside of the solar system. On the other hand, the ITU defines deep space as the region starting at a distance of 2 million kms from the Earth's surface [6]. This definition includes Mars (and Venus), but excludes the Moon, which is at an average distance of 384,000 km from Earth. However, the IETF TIPTOP WG scope also includes the Moon, due to the greater immediacy of the related missions. Hereinafter, the term deep space will comprise the Moon.

### B. Mission Scenarios

We next describe deep-space mission scenarios, including Moon, Mars, celestial body surface, and spacecraft [7].

#### 1) Moon

There are currently six active Moon orbiters, and one active Moon rover. Such numbers are expected to increase in the near future, since NASA and the China National Space Administration (CNSA) plan to build permanently inhabited lunar bases in the 2030s [8]. A variety of devices, including landers, rovers, sensors, and human user devices, will need communication capabilities. When Direct-To-Earth (DTE) communication from such devices is not possible, orbiters are used as relays.

Since 2021, NASA, the European Space Agency (ESA) and the Japan Aerospace eXploration Agency (JAXA) have collaborated on a specification to enable LunaNet, a global network intended to interconnect lunar networks, similar to the terrestrial Internet [9]. Interoperability between the constituent networks, which will be offered by LunaNet Service Providers (LNSPs), is expected to be achieved by means of IP and BP, for different services.

#### 2) Mars

As of the writing, there are two active rovers (Curiosity and Perseverance), and a damaged helicopter (Ingenuity) on Mars' surface. While they may use DTE communication, they are mostly assisted by the Mars Relay Network, a set of four Mars orbiters from NASA and ESA that are used as communication relays. More than 20 missions to Mars are planned in the next two decades, including further landers, rovers, sample return vehicles, and robotic precursors for establishing human habitats on Martian surface. An early form of Mars network (MarsNet) interconnecting surface and relay elements, also connected to Earth networks, has been envisioned by the Interagency Operations Advisory Group (IOAG) [2].

#### 3) Celestial body surface

Communication between devices on a celestial body surface (e.g., the Moon, Mars, an asteroid, etc.) is expected to be enabled by several technologies, including wireless PAN (e.g., IEEE 802.15.4), wireless LAN (i.e., IEEE 802.11), wireless WAN (e.g., 3GPP 4G to 6G), and satellite. In fact, a 4G network was deployed and validated on the lunar surface in March 2025, as part of the Intuitive Machines IM-2 mission. On the other hand, ZigBee (an IoT protocol stack based on IEEE 802.15.4) has been used to communicate between Perseverance and Ingenuity on Mars.

#### 4) Spacecraft

There are currently tens of cruising spacecrafts beyond Earth orbits, ranging from the lunar and Martian environment to the outer solar system and interstellar space. Modern spacecrafts communicate directly with Earth using specialized Consultative Committee for Space Data Systems (CCSDS) link protocols, such as Telecommand (TC), and Telemetry (TM) [2]. Aboard a spacecraft, there are typically several connected computers and sensors, often using real-time communication buses, while IP over LAN variants is gaining adoption.

### C. Deep-Space IoT Use Cases

Current and future IoT communication use cases associated with the described mission environments include the following:

- Telecommand: intended to remotely manage deep-space equipment (e.g., from Earth), such as configuring instruments, modifying vehicle trajectories, etc.
- Telemetry: a service that reports (e.g., periodically) the status of a spacecraft and its components (e.g., to Earth).
- Scientific data collection: typically based on bulk transfer of data from a celestial body. Event detection may trigger multicast messages to coordinate several instruments (e.g., on a planet's surface) to measure phenomena (e.g., space weather anomalies) [7].
- Navigation: a satellite-based system providing location on a celestial body. Orientation is challenging without a global magnetic field (e.g., on Moon or Mars).
- Smart habitat: a self-sustaining module providing conditions for human life in deep space, including radiation protection, air and water recycling, and thermal control. Its systems rely on the information provided by a high number of sensors. The latter need to communicate remotely to enable monitoring from Earth. However, considering long delays and connectivity disruption (see Section III), local autonomy is essential.

## III. Deep Space versus IoT: Networking Characteristics

Deep-space environments present characteristics for network protocols significantly different from those assumed for the classical Internet, yet resemble those encountered in IoT scenarios (Table I). Such characteristics are overviewed next.

### A. High and Variable Delay

Due to the vast distances between celestial bodies, the Round

Trip Time (RTT) in deep space is significantly greater than in the classical Internet. Furthermore, due to orbital dynamics, such distances are highly variable. While the RTT in the classical Internet is assumed to be below 1 second (i.e., the default initial retransmission or probe timeout for TCP and QUIC), the average light-speed RTT between Earth and Moon is 2.56 seconds, and the RTT between Earth and Mars is between approximately 6 and 45 minutes.

Many IoT scenarios also exhibit higher RTT values than classical Internet ones, for several reasons. Firstly, IoT technologies often provide low bit rates, even below 1 kbit/s [10], yielding packet transmission times greater than 1 second. Secondly, many IoT devices rely on simple batteries or energy harvesting solutions as power source. Such devices typically use an energy-saving technique called radio-duty cycling, where they remain in a low-energy state (with the communication interface turned off), for intervals ranging from milliseconds to days. Thirdly, some IoT technologies (e.g., Bluetooth Mesh, LoRaWAN and Sigfox), require a transmission *from* the IoT device to allow an opportunity for sending a message *to* the IoT device. However, the time between two consecutive messages by an IoT device typically spans from seconds to days.

### B. Long Disruptions

In the classical Internet, an end-to-end path is typically composed of a set of links that are simultaneously available. However, in deep space, connectivity with a next hop may not be possible for a relatively long time, due to celestial bodies' rotation and translation, spacecrafts' orbits, and link coverage constraints. For example, Perseverance can use 6 to 8 useful Mars orbiter passes per Martian day, with a visibility window of up to approximately 15 minutes. Moreover, when an orbiter is eclipsed by its celestial body, communication with Earth is not feasible for hours. Furthermore, solar phenomena may produce connectivity blackouts from Earth of hours or days. Such phenomena include solar conjunctions (where a planet is occluded by the Sun), and solar particle events (where the Sun emits a burst of charged particles). While the former are predictable, the latter are not.

IoT devices are also often unavailable for communication. As aforementioned, radio duty-cycled devices remain inactive for relatively long intervals. Furthermore, since IoT networks are typically wireless, sustained disruption may also occur due to changes in the physical environment, such as new obstacles producing unattainable link fading. This phenomenon is exacerbated in multihop topology deployments.

### C. Low Bit Rate

In deep-space missions, communication links primarily use Radio Frequency (RF) signals. Due to the vast distances involved, most bit rates are significantly lower than those typically used on the Internet. For Earth-Moon communication, bit rates are typically in the order of 100 bit/s to 10 kbit/s (uplink, i.e., from Earth) and 1 kbit/s to 100 Mbit/s (downlink). Earth-Mars is even more challenging due to the greater distance, with bit rates in the order of 10 bit/s to 1 kbit/s (uplink) and 10 bit/s to 1 Mbit/s (downlink) [2]. Optical, laser-based links have also been experimented with, offering bit rates in the order of 10 Mbit/s (uplink) and 100 Mbit/s (downlink), as in the Artemis II lunar mission. However, optical links face challenges due to extreme alignment requirements.

Typically, IoT applications have relaxed throughput requirements, thus IoT technologies provide relatively low bit rates. While recent IoT technologies such as 5G RedCap provide up to approximately 100 Mbit/s (intended to support use cases such as video surveillance), most IoT technologies offer a simple connectivity solution for infrequent transmission of small messages. As a result, deep-space bit rates are relatively similar to IoT ones. Bluetooth Low Energy (BLE) and Long Term Evolution for Machines (LTE-M) offer bit rates in the order of 1 Mbit/s. IEEE 802.15.4 and Narrowband IoT (NB-IoT) provide bit rates of 20 kbit/s to 250 kbit/s. LoRaWAN and Sigfox allow minimum bit rates of 250 bit/s and 100 bit/s, respectively.

### D. Asymmetric Bandwidth

In contrast with classical Internet and IoT scenarios, deep-space links are highly asymmetric, up to an asymmetry factor of 67,200 for a given mission [2]. While downlink transmission benefits from very large antennas and a high power capacity on Earth ground stations to amplify the received signals, uplink transmission is challenged by small antennas and low energy availability on deep-space targets. Uplink transmission to lunar environments is further impaired by the Moon's surface radiation, whereby ionized particles from the Sun are absorbed by the lunar regolith and transformed into electromagnetic radiation, contributing to background noise.

A subset of IoT technologies provide asymmetric bit rates for uplink (i.e., from the IoT device) and downlink transmission. Such technologies generally belong to the Low-Power Wide Area Network (LPWAN) family. In Sigfox, in some world regions, the uplink and downlink bit rates are 100 bit/s and 600 bit/s, respectively. In LoRaWAN (EU band), the mandatory uplink bit rate is up to 5 kbit/s, whereas downlink transmission is limited to 250 bit/s in some cases. Nevertheless, many other prominent IoT technologies (e.g., IEEE 802.15.4, BLE, Z-Wave, IEEE 802.11ah, etc.) offer symmetric bit rates. Therefore, deep-space scenarios are significantly more asymmetric.

### E. Computational Constraints

Spacecraft hardware requires protection against solar energetic particles and galactic cosmic rays. For this reason, onboard computers are radiation-hardened and, as a result, they may be resource-constrained, compared to common desktop/laptop computers or even smartphones. In some cases, as in Perseverance, spacecraft computers are based on a single-core processor, clock rates of a few hundred megahertz, and a RAM capacity of a few hundred megabytes, which may require simplified communication protocol implementations.

Typically, IoT devices are also constrained, often to a greater extent. Early IETF work in IoT assumed that a full IP-based protocol stack would need to be supported by so-called Class 1 devices, characterized by ~10 kB of RAM. Modern IoT devices exhibit a RAM capacity of a few hundred kilobytes [10].

TABLE I
MAIN FEATURES OF DEEP-SPACE NETWORKS, IOT NETWORKS, AND THE CLASSICAL INTERNET.

| | Deep Space (Moon) | Deep Space (Mars) | IoT Networks | Classical Internet |
|---|---|---|---|---|
| Medium | Wireless | Wireless | Wireless (typically) | Wired and wireless |
| Topology | Single-hop and multihop | Mostly Multihop | Single-hop and Multihop | Multihop |
| RTT | 2.5 s (average) | 6 to 45 minutes | Milliseconds to days | <1 s (typically) |
| Disruption | Yes (Minutes, hours, days) | | Yes (seconds to days) | No (typically) |
| Bit rate (order of magnitude) | 100 bit/s to 10 kbit/s (RF, uplink), 1 kbit/s to 100 Mbit/s (RF, dwnl.) 10 Mbit/s (optcl., upl.), 100 Mbit/s (optical, downlink) | 10 bit/s to 1 kbit/s (RF, uplink), 10 bit/s to 1 Mbit/s (RF, downlink) | 100 bit/s to 1 kbit/s (LoRaWAN, Sigfox), 10 kbit/s to 100 kbit/s (802.15.4, NB-IoT), 1 Mbit/s (BLE, LTE-M), 100 Mbit/s (5G RedCap) | 100 Mbit/s (4G), 100 Mbit/s to 1 Gbit/s (5G), 1 Gbit/s (Wi-Fi), 1-10 Gbit/s (Ethernet) |
| Asymmetry | High | | None to Low | None to Low |
| Processor | Single-core and multicore | | Single-core | Multicore |
| Clock rate | >100 MHz | | >10 MHz | >1 GHz |
| RAM | From 100s of MB | | From 10 kB | 10s of GB |
| Energy supply | Photovoltaic | Photovoltaic (orbiters), nuclear (rovers) | Small batteries, energy harvesting | Mains power or equivalent |

### F. Energy Constraints

Deep-space equipment has a constrained energy supply. Moon vehicles and Mars orbiters typically use photovoltaic energy. As distance from the Sun increases beyond Jupiter, nuclear generators become the preferred spacecraft power source. Such generators are also used by current Mars rovers, since dust storms may render solar panels useless. However, the available power (e.g., 100 W for Perseverance) needs to be shared by all operations of the spacecraft, limiting communication performance and connectivity opportunities.

Energy availability is also a fundamental problem for IoT devices, and one of the main reasons why IoT-specific communication protocols have been designed. To avoid dependency on cabled power solutions, many IoT devices run on a simple battery (e.g., of 230 mAh) or operate based on energy harvesting solutions, which are intrinsically limited.

## IV. PROTOCOL ARCHITECTURES FOR DEEP SPACE

This section overviews the BP-based and IP-based protocol architectures for deep-space environments, and presents an IP-based IoT protocol stack suitable for such scenarios.

### A. BP-Based Protocol Architecture

In the early 2000s, the concept of an Interplanetary Internet was envisioned. However, the challenging features of deep-space communication, especially extreme delays and connectivity disruption, were considered unsuitable for TCP/IP, which had been designed for the terrestrial Internet. Features such as TCP's three-way handshake and congestion control based on terrestrial RTT-scale receiver feedback, are unfit for deep-space networks. This problem motivated the design of the BP-based architecture (Fig. 2.a) [3].

When BP is used, an application-layer message is encapsulated as the payload of a data unit called a *bundle*, which is carried atop the transport layer (possibly over various lower-layer hops, e.g., a full Internet path) until the bundle reaches the next *bundle node*. The latter stores the bundle until a link with the next bundle node (e.g., over deep space) becomes available. Then, the bundle node relays the bundle, possibly over a different set of transport- and lower-layer protocols. This process repeats iteratively until the bundle reaches the destination. The BP-based architecture defers reliability and congestion control to the lower layers, following a BP-layer hop-by-hop approach.

### B. IP-Based Protocol Architecture for Deep Space

The renewed interest in space exploration has accelerated plans to deploy IP-based networks on or near celestial bodies. Using IP to interconnect such networks is a natural solution that furthermore simplifies the protocol stack (Figs. 2a and 2.b). Note that intermediate nodes can function as simple IP routers.

Fig. 3 (left half) illustrates the IP-based protocol stack for deep-space environments first considered by the IETF TIPTOP WG [5]. IP provides the core of the protocol stack, as in the terrestrial Internet, albeit enhanced with persistent packet storage capabilities. Below IP, various lower-layer technologies can be used, depending on the physical environment. Atop IP, UDP is the main classic transport-layer protocol assumed, as it is lightweight and time-independent, whereas TCP is discarded due to its aforementioned issues in deep space. Thus, UDP-based protocols are candidates for use in deep space. QUIC is the main one, as it can outperform TCP, its adoption is increasing, and the number of QUIC-supported upper-layer protocols and applications is also expanding, including HTTP/3, Domain Name System (DNS), media, and NETCONF (a network management solution), among others. QUIC comprises modern TCP functionality, TLS 1.3, and further improvements intended to efficiently provide end-to-end reliability and security. A significant QUIC feature is that, compared with using TCP and TLS separately, it reduces the connection establishment latency (down to a 0-RTT overhead in some cases). This is crucial, considering the extreme delays in deep-space environments. However, since QUIC was designed for the classical Internet, it requires adaptation for suitable operation in such environments [12].

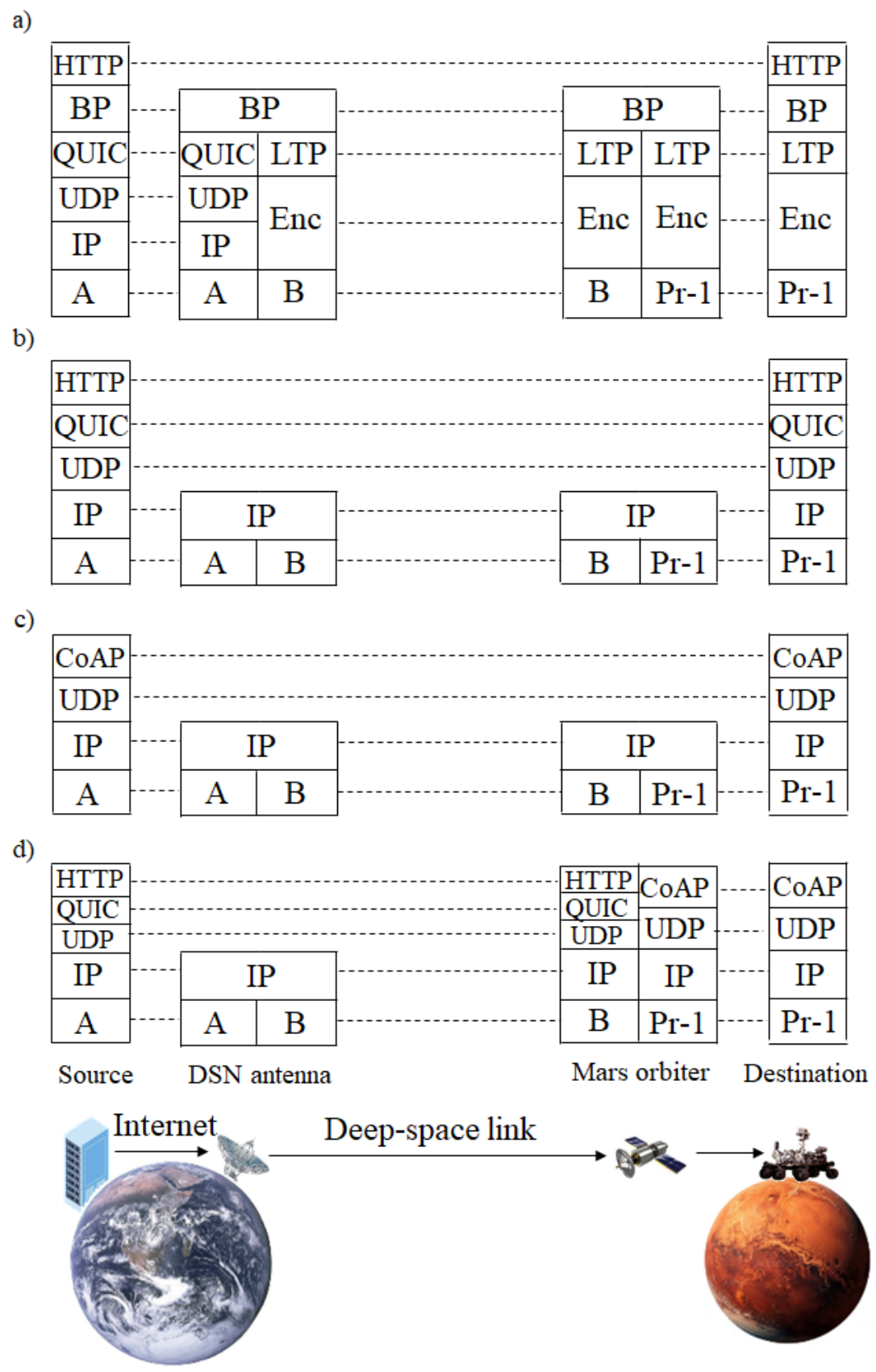

Fig. 2. Earth-Mars communication: a) BP-based, b) IP-based, c) IoT IP-based, d) HTTP-CoAP proxy scenario. A represents a terrestrial technology. B denotes a CCSDS deep-space technology, such as TC. Proximity-1 (Pr-1) is a CCSDS technology for Mars orbiter to surface communication. Licklider Transmission Protocol (LTP) is a specialized transport-layer protocol for deep space, often used below BP. Enc stands for Encapsulation. Deep Space Network (DSN) antennas are used by NASA to communicate with spacecrafts beyond Earth's orbit. Note: SCHC may be used in scenarios c) and d).

### C. IP-Based IoT Protocol Architecture for Deep Space

Given the opportunity to use IP in deep space, and the similarity between such environments and IoT ones, there is a clear rationale to consider IP-based IoT protocols for the Interplanetary Internet.

Fig. 3 (right half) illustrates an IP-based IoT protocol stack for deep space. IP-based IoT protocol stacks commonly use an adaptation layer below IP. Typically, the adaptation layer provides two main functions: packet header compression, and packet fragmentation. The IETF has standardized two main adaptation layer frameworks for IoT networks: IPv6 over Low power WPAN (6LoWPAN), initially designed to support IPv6 over IEEE 802.15.4 networks, and Static Context Header Compression and fragmentation (SCHC), originally developed to support IPv6 over LPWANs [14]. While 6LoWPAN has only been defined to compress IPv6 and UDP packet headers, SCHC is available for a wider range of protocols, and provides a greater header compression performance [14]. For this reason, SCHC is preferred over 6LoWPAN in order to save energy and bandwidth in deep-space scenarios. Adaptation-layer fragmentation is needed when the underlying link layer uses small frames and it does not support its own fragmentation mechanisms, as in IEEE 802.15.4, LoRaWAN and Sigfox.

| HTTP/3 | Media | DNS | NETCONF | ··· | ··· | LwM2M | DNS | CORECONF | EDHOC | ··· |
|---|---|---|---|---|---|---|---|---|---|---|
| QUIC | | | | | ··· | CoAP | | | | |
| UDP | | | | | | | | | | |
| IP | | | | | SCHC | | | | | |
| Link and physical layers | | | | | | | | | | |

Fig. 3. IP-based protocol stack for deep space. The light-background stack (right) is based on IoT protocols.

Similarly to the TIPTOP WG architecture, the efforts from the IETF for end-to-end communication in IoT use cases primarily focus on UDP-based functionality. The Constrained Application Protocol (CoAP) is an application-layer protocol with embedded transport-layer mechanisms that was originally designed to operate atop UDP for resource-constrained environments [11]. In addition to supporting IoT applications, there is a growing ecosystem based on CoAP, which carries LwM2M (a device lifecycle management solution used mainly in cellular networks), CORECONF (a low-overhead network management solution), EDHOC (a lightweight authenticated key exchange), and DNS.

## V. IoT Protocols in Deep Space

This section discusses SCHC and CoAP in the context of deep space.

### A. SCHC in Deep Space

SCHC was designed assuming a static header compression context shared by the compressor and the decompressor, for a multiyear IoT device battery lifetime. Not requiring context updates is especially appropriate for deep space.

Considering the connectivity time windows, as well as bandwidth and energy constraints in deep space, using SCHC can provide significant performance improvement. Currently, SCHC can compress IPv6, UDP, CoAP and ICMPv6 packet headers. It is also possible to perform joint header compression of a set of protocol headers. Using SCHC, a 52-byte IPv6/UDP/CoAP joint header can be compressed down to a 2-byte format, increasing the effective network capacity. For example, using SCHC to compress IPv6/UDP/CoAP packets carrying a 27-byte CoAP payload allows to duplicate the network capacity over CCSDS TC. Greater improvement is possible for larger CoAP header sizes.

As of the writing, a preliminary design for SCHC to compress QUIC headers has been proposed. We suggest that SCHC-based HTTP header compression would further increase performance. Another area that has recently been explored is SCHC-based compression of structured payloads (e.g., JSON-formatted sensor readings).

SCHC fragmentation is not needed over CCSDS link protocols, as they provide their own fragmentation mechanisms.

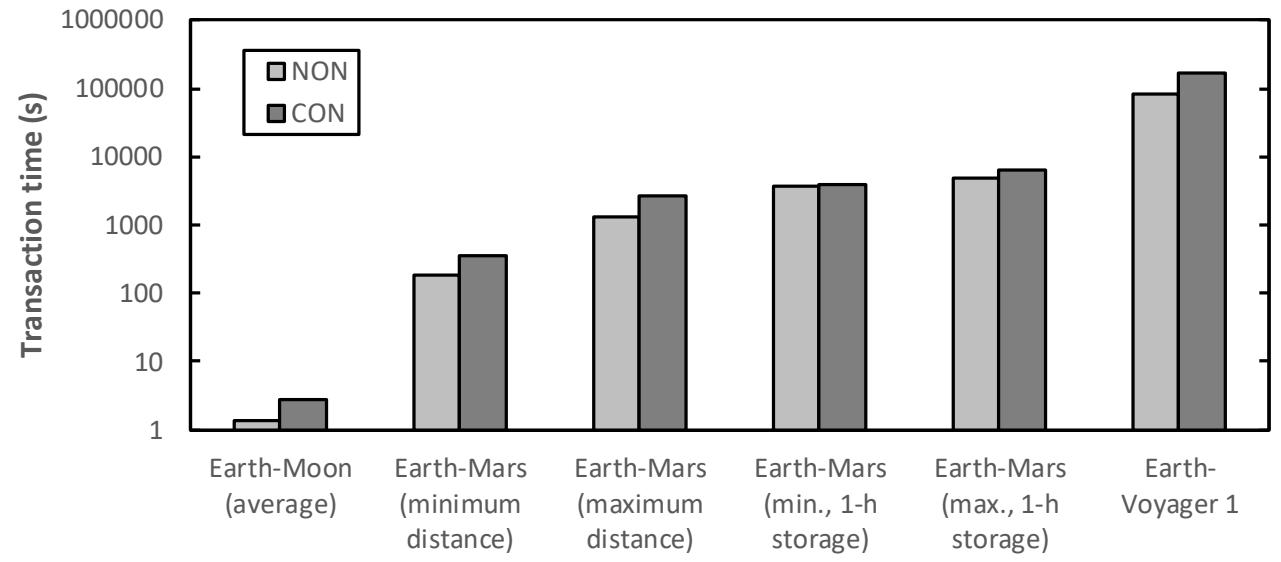

Fig. 4. Transaction times for CoAP NON and CON messages for Earth-Moon, Earth-Mars and Earth-Voyager 1 scenarios. The latter provides an upper bound on light speed transaction time. A packet storage delay of 1 hour has been considered for the Earth-Mars scenario. Earth-Voyager 1 does not use a relay.

### B. *CoAP in Deep Space*

CoAP provides lightweight operation (with a 4-byte header size excluding optional fields), and significant flexibility. Similarly to HTTP, CoAP follows a client-server model, whereby a client sends requests aiming to manipulate resources managed by a server, which provides responses to the clients.

CoAP is composed of two sublayers. The upper one handles requests and responses, whereas the lower one provides transport-layer functionality to messages, since CoAP runs by default atop UDP. Such functionality includes per-message optional reliability, which is achieved by using acknowledgments (ACKs), stop-and-wait operation by default, and Retransmission TimeOut (RTO)-based message retransmission with binary exponential back-off for congestion control. CoAP was also provided with support over reliable transports, such as TCP, which however are not recommended for deep space, as previously described. Hereinafter, we will refer to CoAP over UDP (Fig. 2.c).

CoAP supports caching and proxying. When combined, such features allow an intermediary (e.g., a proxy at a Mars orbiter) to cache responses from origin servers (e.g., sensors on Mars' surface), so that subsequent requests (e.g., from Earth) on the same resources can be serviced directly by the proxy if they are recent enough, thus saving valuable resources. Furthermore, CoAP's upper layer was designed to allow easy mapping with HTTP messages through a protocol translation proxy (e.g., the Mars orbiter in Fig. 2.d). CoAP also offers an optional mechanism called Observe, which provides updates on the state of a resource to interested clients, avoiding the need for repeated requests.

Since UDP does not provide data unit segmentation, CoAP supports reliable blockwise transfer mechanisms, which allow sending large application-layer data units. One such mechanism is based on a negative ACK (NACK)-oriented scheme, which is efficient and suitable for deep-space environments.

CoAP also offers message aggregation, which allows exploiting connectivity opportunities while reducing protocol overhead. For example, a CoAP endpoint (e.g., a Mars orbiter proxy) can produce an aggregate CoAP message composed of several individual messages (e.g., carrying data from a Mars rover) while there is no connectivity with the next hop (e.g., Earth).

A distinctive CoAP feature, not supported by unicast protocols like QUIC, is group communication, which can enable deep-space multicast use cases efficiently.

CoAP was originally designed to be secured by means of Datagram TLS (DTLS). However, DTLS involves handshakes between the involved endpoints that would incur a high delay in deep-space scenarios. Another security approach for CoAP is Object Security for Constrained RESTful Environments (OSCORE), which allows to protect CoAP messages with encryption and integrity end to end, even in the presence of untrusted proxies in the path. In OSCORE, it is possible to avoid handshakes if the security context is preshared between the involved endpoints. For these reasons, OSCORE is recommended to secure CoAP in deep-space scenarios.

## VI. CoAP Adaptation for Deep Space

Despite its proper features for deep space, CoAP requires adaptation for such scenarios [13].

### A. *Parameter Settings*

The initial CoAP RTO is chosen randomly between 2 and 3 seconds. The random RTO component is intended to avoid synchronization between neighboring nodes performing simultaneous retransmissions, eventually leading to repeated packet collisions. Since the greatest light-speed RTT for Earth-Moon communication is 2.71 seconds, the initial CoAP RTO needs to be set to at least the same value, to avoid spurious retransmissions. However, for Earth-Mars communication, the initial CoAP RTO needs to be increased to at least the specific light-speed RTT for the distance between the two planets at the moment of communication. Such value will be between 365 seconds and 2642 seconds. Similar time increases apply to various other CoAP timer-based parameters, including the amount of time during which a message identifier cannot be reused. Further delays (e.g., due to packet store and forward) need to be considered. To avoid incurring undue additional delay, the random RTO component must be disabled in the Earth-Mars scenario, since the greater RTO scale renders the probability of transmission synchronization negligible.

Fig. 4 illustrates the transaction time for CoAP non-confirmable (NON) and confirmable (CON) messages in different scenarios, assuming no packet loss and appropriately set CoAP parameters. For CON messages, transaction time includes reception of the corresponding ACK. Note that spacecraft packet storage delays in the order of hours are common, and they often represent the main contributor to total communication delay.

Another important CoAP parameter is the maximum number of message retries, equal to 4 by default. This parameter needs to be configured considering the trade-off between reliability, latency, bandwidth and energy consumption. Reducing the value of this parameter may be suitable for the Earth-Mars scenario, based on its greater RTT scale, the more limited connectivity opportunities, and the robustness of CCSDS link protocols, which use advanced Forward Error Correction (FEC).

### B. Congestion Control

The binary exponential back-off scheme used in CoAP for CON messages was designed to parallel TCP's similar congestion control component, with the aim to avoid risk of congestion collapse. However, TCP's congestion control is based on a closed-loop approach that relies on RTT-scale receiver feedback (based on the presence/absence of ACKs, or explicit congestion-related indications). While such approach is suitable for the terrestrial Internet, it is inappropriate in deep space, whereby congestion signals would arrive too late for the sender to react. For this reason, and leveraging a priori knowledge of path features, open-loop congestion control can be used [12]. A CoAP sender can transmit packets at appropriate times and rates. Another approach is relying solely on flow control jointly with buffering, both appropriately dimensioned based on precomputed delays and disruption interval durations.

### C. Path Asymmetry

CoAP can be tuned to handle the path asymmetry of a deep-space scenario, where ACKs via a slow reverse path may represent a bottleneck to data transmission via the forward path. In contrast with QUIC or TCP, CoAP offers NON messages, which do not elicit ACKs. A balance between NON and CON messages may be selected considering the reliability requirements of the involved applications. However, while CoAP allows going beyond the default stop-and-wait behavior for CON messages, an ACK is required for every CON message. Using cumulative ACKs in CoAP is a promising future work item.

## VII. Conclusions

Ongoing IETF standardization and related research activities provide the cornerstone for creating an IP-based Interplanetary Internet. This new paradigm facilitates the use of IoT protocols in such a scenario. With proper adaptation, IoT protocols can exploit the similarities between IoT and deep-space environments for efficient operation in the latter.

## Acknowledgment

Carles Gomez has been supported in part by the Spanish Govt.'s Ministerio de Ciencia, Innovación y Universidades MCIU/AEI/10.13039/501100011033/FEDER/UE through project PID2023-146378NB-I00, and the Estancias de Movilidad en Centros Extranjeros de Enseñanza Superior e Investigación PRX24/00397 grant.

## References

[1] L. Giraldi, "A Comprehensive Overview of the Growing Global Space Economy: From Government Leadership to Commercial Opportunities and Ecosystem Development", In Space Economy SEWA - Start, Evolve, or Walk Away. International Series in Advanced Management Studies, Springer, 2025.

[2] "Volume 1. The Future Mars Communications Architecture", Report of the Interagency Operations Advisory Group. Mars and Beyond Communications Architecture Working Group, Feb. 2022. https://ioag.org/wp-content/uploads/gravity_forms/6-1a23b5e730ab942f4b44183168a5bb93/2025/04/MBC-architecture-report-final-version-PDF.pdf (Accessed on Aug. 15, 2026.)

[3] V. Cerf, S. Burleigh, L. Torgerson, R. Durst, K. Scott, K. Fall, H. Weiss, "Delay-Tolerant Networking Architecture", RFC 4838, Apr. 2007. https://www.rfc-editor.org/rfc/rfc4838 (Accessed on Aug. 15, 2026.)

[4] K. Fall, "A delay-tolerant network architecture for challenged internets", in proc. of SIGCOMM'03, Karlsruhe, Germany, Aug. 2003.

[5] M. Blanchet, W. Eddy, T. Li, "An Architecture for IP in Deep Space", IETF Internet Draft, https://datatracker.ietf.org/doc/draft-ietf-tiptop-ip-architecture (work in progress). (Accessed on Aug. 15, 2026.)

[6] "Handbook on Space Research Communication", International Telecommunication Union - Radiocommunication Sector, Edition of 2026, ITU Publications, 2026.

[7] M. Blanchet, W. Eddy, M. Eubanks, "IP in Deep Space: Key Characteristics, Use Cases and Requirements", IETF Internet Draft, https://datatracker.ietf.org/doc/draft-ietf-tiptop-usecase (work in progress). (Accessed on Aug. 15, 2026.)

[8] Y. Xia, Y. Gao, W. Han, X. Li, C. Zhou, Y. Zhou, and L. Ding, "Lunar base infrastructure construction: Challenges and future directions", Automation in Construction, Vol. 176, Art. No. 106251, May 2025.

[9] "LunaNet Interoperability Specification Document", version 5, Jan. 2025. https://www.nasa.gov/wp-content/uploads/2025/02/lunanet-interoperability-specification-v5-baseline.pdf?emrc=606f95 (Accessed on Aug. 15, 2026.)

[10] C. Bormann, M. Ersue, A. Keränen, C. Gomez, "Terminology for Constrained-Node Networks", IETF Internet Draft, draft-ietf-iotops-7228bis (work in progress). (Accessed on Aug. 15, 2026.)

[11] C. Bormann, A. Castellani, Z. Shelby, "CoAP: an Application Protocol for Billions of Tiny Internet Nodes", IEEE Internet Computing, Vol. 16, No. 2, Mar./Apr. 2012, pp. 62-67.

[12] M. Blanchet, W. Eddy, "QUIC Profile for Deep Space", IETF Internet Draft, https://datatracker.ietf.org/doc/draft-many-tiptop-quic-profile/ (work in progress). (Accessed on Aug. 15, 2026.)

[13] C. Gomez, S. Aguilar, "CoAP in Space", IETF Internet Draft, https://datatracker.ietf.org/doc/draft-gomez-tiptop-coap (work in progress). (Accessed on Aug. 15, 2026)

[14] C. Gomez, A. Minaburo, L. Toutain, D. Barthel, J.C. Zuniga, "IPv6 over LPWANs: Connecting low power wide area networks to the Internet (of Things)", IEEE Wireless Comm.., Vol. 27, No. 1, 206-213, Feb. 2020.

## Biographies

Carles Gomez received his Ph.D. degree from Universitat Politècnica de Catalunya in 2007. He is a Full Professor at the same university. He is a co-author of numerous technical contributions (papers, IETF RFCs, and books). He is a co-chair of the IETF 6Lo working group. His research interests focus mainly on IoT and deep-space communications.

Jon Crowcroft has been the Marconi Professor of Communications Systems in the Computer Laboratory, University of Cambridge since 2001. He has worked in the area of Internet support for multimedia communications for over 40 years. Three main topics of interest have been scalable multicast routing, practical approaches to traffic management, and the design of deployable end-to-end protocols.